\documentclass[12pt,preprint]{aastex} 

\shorttitle{Discovery of HMgNC}
\shortauthors{Cabezas et al.}

\begin{document}


\title{Laboratory and Astronomical Discovery of HydroMagnesium Isocyanide
\thanks{This work was based on observations carried out with the 
IRAM 30-meter telescope. IRAM is supported by INSU/CNRS (France), 
MPG (Germany) and IGN (Spain)}}

\author{C. Cabezas}
\affil{Grupo de Espectroscopia Molecular (GEM), Unidad Asociada CSIC, Edificio Quifima. 
Laboratorios de Espectroscopia y Bioespectroscopia. Universidad de Valladolid, 47005 Valladolid, Spain}
\email{ccabezas@qf.uva.es}

\author{J. Cernicharo}
\affil{Deparment of Astrophysics, CAB. INTA-CSIC. Crta Torrej\'on-Ajalvir Km\,4. 28850 Torrej\'on
de Ardoz, Madrid, Spain}

\and
\author{J.L. Alonso}
\affil{Grupo de Espectroscopia Molecular (GEM), Unidad Asociada CSIC, Edificio Quifima. 
Laboratorios de Espectroscopia y Bioespectroscopia. Universidad de Valladolid, 47005 Valladolid, Spain}

\and
\author{M. Ag\'undez}
\affil{University of Bordeaux, LAB, UMR 5804, F-33270 Floirac, France} 

\and
\author{S. Mata}
\affil{Grupo de Espectroscopia Molecular (GEM), Unidad Asociada CSIC, Edificio Quifima. 
Laboratorios de Espectroscopia y Bioespectroscopia. Universidad de Valladolid, 47005 Valladolid, Spain}

\and
\author{M. Gu\'elin}
\affil{Institut de Radioastronomie Millim\'etrique, 300 rue de la Piscine, 38406 Saint Martin d'H\'eres, France}

\and
\author{I. Pe\~na}
\affil{Grupo de Espectroscopia Molecular (GEM), Unidad Asociada CSIC, Edificio Quifima. 
Laboratorios de Espectroscopia y Bioespectroscopia. Universidad de Valladolid, 47005 Valladolid, Spain}

\begin{abstract}
We report on the detection of hydromagnesium isocyanide, HMgNC, in the laboratory and in the
carbon rich evolved star IRC+10216. The J=1-0 and J=2-1 lines were observed in our microwave
laboratory equipment in Valladolid with a spectral accuracy of 3\,KHz. The hyperfine structure 
produced by the Nitrogen atom was resolved for both transitions. The derived rotational constants
from the laboratory data are $B_0$=5481.4333(6)\,MHz, $D_0$=2.90(8)\,KHz, and $eQq(N)$=-2.200(2)\,MHz. 

The predicted frequencies 
for the rotational transitions of HMgNC in the millimeter domain have an accuracy of 0.2-0.7\,MHz.
Four rotational lines of this species, J=8-7, J=10-9, J=12-11 and J=13-12, have been detected towards
IRC+10216. The differences between observed and calculated
frequencies are $<$0.5\,MHz. The rotational constants derived from space frequencies
are $B_0$=5481.49(3)\,MHz and $D_0$=3.2(1)\,KHz, i.e., identical to the laboratory ones.
A merged fit to the laboratory and space frequencies provides 
$B_0$=5481.4336(4)\,MHz and $D_0$=2.94(5)\,KHz.

We have derived a column density for HMgNC of (6$\pm$2)$\times$10$^{11}$\,cm$^{-2}$. From the observed 
line profiles the molecule have to be produced produced in the layer where other metal-isocyanides 
have been already found in this source. The abundance ratio between MgNC and its hydrogenated 
variety, HMgNC, is $\simeq$20.
\end{abstract}


\vspace{0.5cm}
To appear in the Astrophysical Journal September 2013
\vspace{0.5cm}

\section{Introduction}
Metal-bearing molecules,  NaCl, KCl, AlF, and NaF, were detected 30 years ago in the circumstellar
envelope (CSE) of the carbon-rich star envelope IRC+10216 by \citet{Cernicharo1987}. This species were
predicted to be produced under thermodynamical chemical equilibrium near the photosphere
of the star \citep{Tsuji1973}. Just a few months before these detections \citet{Guelin1986}
reported on the presence of a new free radical in IRC+10216. Several Silicon and Sulfur 
bearing candidates were proposed
at that time, but the line carrier was definitively identified as MgNC only years
latter from laboratory measurements \citet{Kawaguchi1993}.
The emission of magnesium isocyanide was mapped in IRC+10216 with the IRAM Plateau de Bure 
Interferometer by Gu\'elin, Lucas and Cernicharo (1993). Unlike metal-halogen species
that are formed near the star photosphere, MgNC was found in a thin shell 15''
in radius where many  reactive species such as the carbon chain radicals C$_5$H, C$_6$H, C$_7$H, and
C$_8 $H, are also detected  \citep{Cernicharo1986a,Cernicharo1986b,Guelin1987,Cernicharo1987a,
Cernicharo1987b,Cernicharo1996,Guelin1997}.
Just one year after the identification of MgNC, sodium cyanide was also detected in IRC+10216
by \citet{Turner1994}. MgNC has been also found towards the more evolved carbon star CRL2688
\citep{Highberger2003}.

It is worth noting that metals such
as Na, K, Ca, Fe, Cr, and/or their cations are found in the gas phase in IRC+10216 
\citep{Mauron2010} pointing towards a rich metal chemistry in the outer circumstellar envelope (CSE).
Since these early works several more metal cyanides or isocyanides
have been found in CSEs. MgCN, AlNC, SiCN, SiNC, KCN, and FeCN have been identified
in IRC+10216 after their mm-wave rotational spectrum has been characterized in the spectroscopic laboratory
\citep{Ziurys1995,Ziurys2002,Guelin2000,Guelin2004,Pulliam2010,Zack2011}.
Many other have been searched for in the same way without success. 
We note that TiO and TiO$_2$ have recently been detected in the Oxygen-rich CSE VyCMa \citep{Kaminski2013}.

While some metal-bearing species containing Al, Na or K are stable closed shell molecules 
that mostly form in the hot part of the envelope, close to the star, other, containing Mg, Si or Fe, are open 
shell radicals that react even at low temperature with neutral molecules or atoms and may be formed in the outer envelope.
MgNC and MgCN remain the only Mg-bearing species detected in space so far. The chemistry of these molecules could be based 
on the reaction of Mg$^+$ with other molecules formed in the CSE \citep{Petrie1996,Dunbar2002}. The detection of additional Mg-bearing 
species could  allow a detailed chemical analysis of the reactions leading to the formation of these molecules and to help 
in discriminating between gas phase chemical paths or dust grain surface reactions.
The adjunction of an hydrogen atom to the radicals MgNC and SiCN yields stable closed shell molecules that may be thought 
to be more abundant. Despite the characterization of their mm spectra in the laboratory \citep{Sanz2002}, HSiCN and HSiNC 
have not been detected so far in space. 

The most abundant metal cyanide being MgNC, it was interesting to see if HMgNC is present.
In this Letter we report on the detection in the laboratory and in space of Hydrogenated
Magnesium Isocyanide, HMgNC and we perform detailed chemical models to analyse the formation
mechanisms of these metal-bearing species.

\section{Laboratory Characterization of HMgNC}
The HMgNC spectrum was measured using a laser ablation molecular beam Fourier transform microwave 
(LA-MB-FTMW) spectrometer at the University of Valladolid, which operates in the 4-26 GHz 
frequency range and is described elsewhere \citep{Alonso2009}. HMgNC was created by 
laser ablation of magnesium rods in the throat of a pulsed supersonic expansion of highly 
diluted ehtylcyanide (0.2\%) in Neon (15 bars stagnation pressure) using a pulsed Nd:YAG-laser 
($\lambda$= 355\,nm, $\simeq$20 mJ\,pulse$^{-1}$) focused on the rod for ablation at a repetition 
rate of 2\,Hz. The rod was continuously rotated and translated in order to minimizing the problem 
of shot-to-shot fluctuation in the amount of the desorbed material. Briefly, the sequence of 
an experimental cycle starts with a gas pulse of the mixing carrier gas (typically 450 $\mu$s). 
After an adequate delay, a laser pulse hits the metal rod producing the vaporization of the solid 
and the chemical reaction in the precursors mixture \citep{Cabezas2012}. Immediately, the resulting 
products are supersonically expanded between the two mirrors of the Fabry-P\`erot resonator and 
then a microwave pulse ($\simeq$0.3\,$\mu$s) is applied, producing the macroscopic polarization of the 
species in the jet. Once the excitation ceases, molecular relaxation gives rise to a transient 
emission signal (free induction decay) at microwave frequencies, which is captured in the time 
domain. Its Fourier transformation to the frequency domain yields the rotational transitions that 
appear as Doppler doublets, because the supersonic jet travels parallel to the resonator axis 
(see Figure 1). The molecular rest frequencies are calculated as the arithmetic mean of the 
Doppler doublets, and are obtained with accuracy better than 3\,kHz. 

To optimize the experimental conditions for HMgNC, we tested the J= 1-0 transition for 
the radical MgNC \citep{Kawaguchi1993,Walker1998} due to the similarities between both molecules. 
A previous theoretical work available in the literature predicts the rotational constant for HMgNC 
molecule around $B_0\simeq$5438.2\,MHz \citep{Gronowski2013}. We have carried 
out ab initio calculations at MP2/6-311++G(d,p) level of theory to predict another 
parameter relevant to our experiment, the nitrogen (I=1) quadrupole coupling constant for which we
obtain $eQq$=-1.95\,MHz. Frequency scans were conducted to search for the J= 1-0 transition around the 
11 GHz range. We finally found a set, composed of three lines (Figure 1) in the region of 10962 MHz 
for which frequency separation was very similar to the expected one, taking into consideration 
the nitrogen eQq value. No other fine or hyperfine structure arising from spin-rotation interactions 
was observed, indicating that the observed spectrum arises from a closed-shell molecule. The next J 
transition was then observed, leading to the unambiguous assignment of HMgNC as a linear molecule, 
since five new hyperfine components were measured in the 21.9\,GHz region. Table 1 lists J= 1-0 and 
J= 2-1 rotational transitions for HMgNC labeled by quantum number F, where F=J+I. 
To analyze them we have used a Hamiltonian of the following form: $H = H_R + H_Q$ where $H_R$ contains 
rotational and centrifugal distortion parameters while $H_Q$ the quadrupole coupling interactions. 
We obtain $B_0$=5481.4333(6)\,MHz, $D_0$=2.90(8)\,KHz, and $eQq(N)$=-2.200(2)\,MHz. The
standard deviation of the fit is 1.6 KHz.

\begin{deluxetable}{rrrrrrr} 
\tablecaption{Laboratory and Space Frequencies for the observed transitions of HMgNC
\label{tab:lines}}
\tablecolumns{7}
\tablehead{
 J$_u$ & F$_u$ & J$_l$ & F$_l$ & $\nu$(MHz) & Unc(MHz) & $\nu_o$-$\nu_c$(KHz)}
\startdata
  1 & 1 &  0 & 1 &   10962.305 &  0.002   &    0.2\\
  1 & 2 &  0 & 1 &   10962.966 &  0.002   &    0.3\\
  1 & 0 &  0 & 1 &   10963.954 &  0.002   &   -1.8\\
  2 & 2 &  1 & 2 &   21924.979 &  0.002   &   -1.2\\
  2 & 1 &  1 & 0 &   21925.089 &  0.002   &   -1.5\\
  2 & 2 &  1 & 1 &   21925.643 &  0.002   &    2.5\\
  2 & 3 &  1 & 2 &   21925.688 &  0.002   &    0.5\\
  2 & 1 &  1 & 1 &   21926.741 &  0.002   &    0.4\\
  3 &   &  2 &   &   32888.284 &  0.003   & predicted\\
  4 &   &  3 &   &   43850.716 &  0.009   & predicted\\
  5 &   &  4 &   &   54812.866 &  0.019   & predicted\\
  6 &   &  5 &   &   65774.663 &  0.035   & predicted\\
  7 &   &  6 &   &   76736.036 &  0.058   & predicted\\
  8 &   &  7 &   &   87697.500 &  0.400   &  584.5\\
 10 &   &  9 &   &  109616.700 &  0.400   & -210.4\\
 12 &   & 11 &   &  131533.900 &  1.000   & -182.6\\
 13 &   & 12 &   &  142490.900 &  1.000   & -533.8\\
 14 &   & 13 &   &  153447.868 &  0.489   & predicted\\
 15 &   & 14 &   &  164403.313 &  0.603   & predicted\\
 16 &   & 15 &   &  175357.701 &  0.733   & predicted\\
 17 &   & 16 &   &  186310.959 &  0.881   & predicted\\
 18 &   & 17 &   &  197263.018 &  1.048   & predicted\\
 19 &   & 18 &   &  208213.806 &  1.233   & predicted\\
\enddata
\tablecomments{The J=1-0 and J=2-1 lines have been observed in the laboratory
with an accuracy of 2 KHz. The lines J=8-7, 10-9, 12-11 and 13-12
have been observed in IRC+10216. Predictions
obtained with the derived rotational constants are provided for lines
with E$_{upp}<$100\,K.}
\end{deluxetable}

\begin{figure}
\includegraphics[angle=0,scale=1.2]{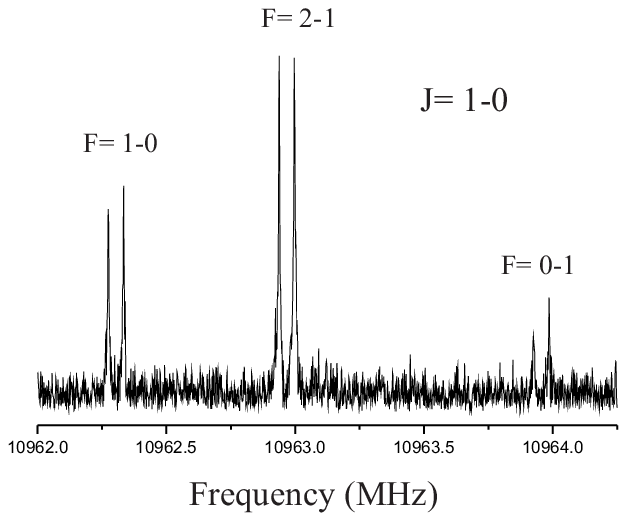}
\caption{J=1-0 rotational transition of HMgNC, observed in this work near to 11 GHz. The nuclear quadrupole coupling 
hyperfine structure is clearly resolved. This spectrum represents one scan at steps of 0.3 MHz with 600 averages 
phase-coherently coadded in each point.}
\label{fig:laboratory}
\end{figure}

\section{Astronomical Observations and identification of HMgNC}
The observations were performed 
at the IRAM 30m telescope at Pico Veleta (Spain). 
Many different runs have been used for the present observations covering the period 1990-2010.
Hence, different receivers and spectrometers have been used. The spectral resolution has
been always 1 MHz. Receiver temperatures were around 150 K in the early observations and have
been improving with time down to 50-60 K in the last observing runs.
All the observations were performed using the Wobbler Switching mode which produces remarkable flat
baselines. Pointing errors were always within 3$''$. The 30m beam size at the observing frequencies
ranges from 29'' for the J=8-7 transition to the 17'' of the J=13-12 line.
The spectra were calibrated in antenna temperature corrected for atmospheric 
attenuation using the ATM package \citep{Cernicharo1985,Pardo2001}. Most of the 3\,mm data pertain
to a line survey that will be published elsewhere (Cernicharo et al., 2013, in preparation).

\begin{figure}
\includegraphics[angle=0,scale=.75]{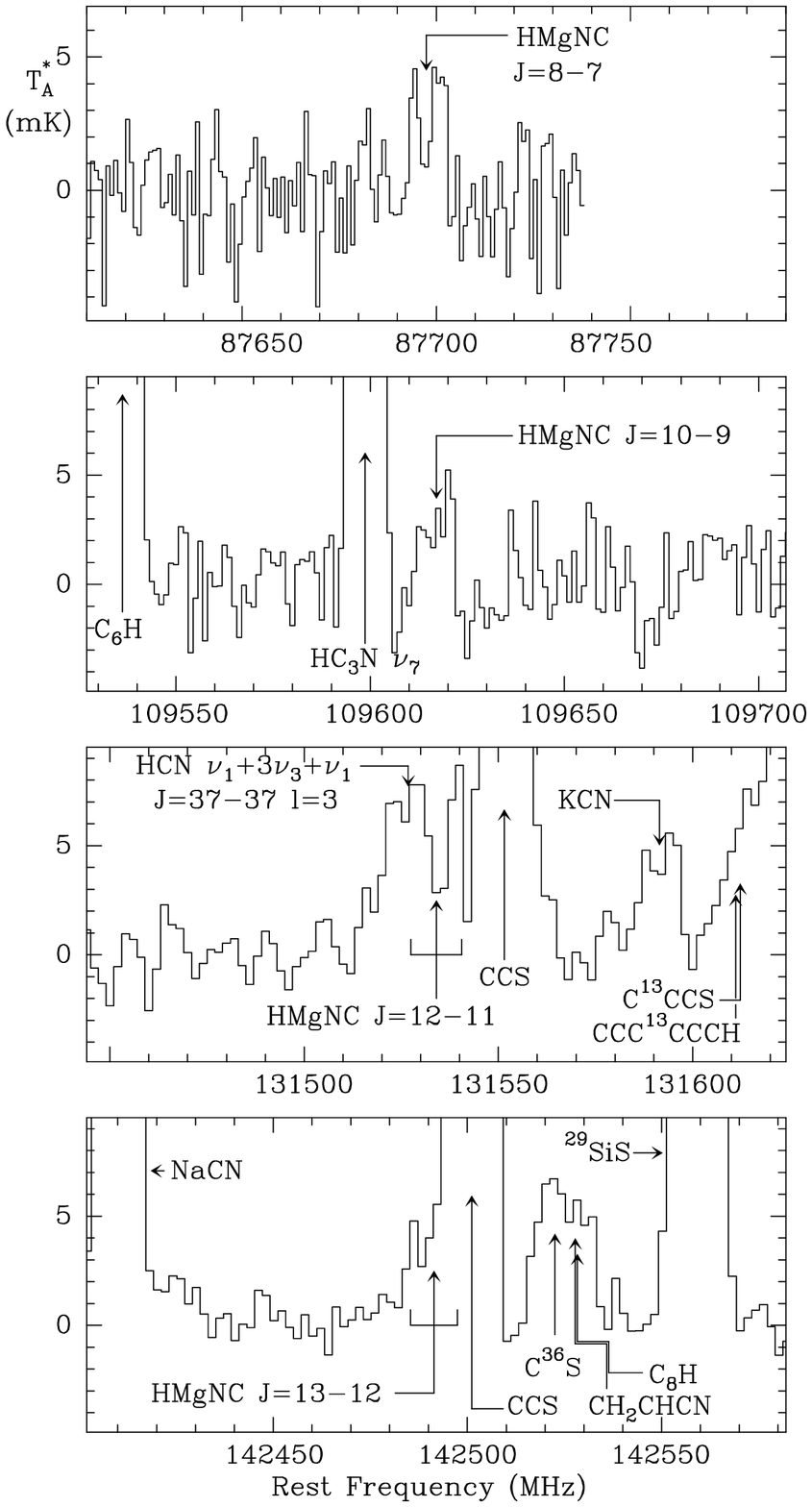}
\caption{Observed transitions of HMgNC towards IRC+10216. The horizontal scale corresponds to
the rest frequency assuming a V$_{LSR}$ of -26.5 km\,s$^{-1}$. The intensity scale is antenna
temperature in mK corrected for atmospheric attenuation and telescopes losses. The J=9-8 line
is fully blended with a line of H$_2$C$_4$ and it is not shown. The J=12-11 is partially blended
with a narrow and weak transition of HCN coming from the dust formation zone \citep{Cernicharo2011}.
The J=13-12 line is also partially blended with a line of CCS but 80\% of the frequency coverage
of the HMgNC line is free of contamination and appears clearly detected. This spectrum
has a sensitivity of 0.6 mK (1 $\sigma$). The whole frequency coverage of the J=12-11 and J=13-12
transitions is indicated.
}
\label{fig:astronomical}
\end{figure}

The rotational constants of HMgNC determined in the laboratory have been implemented in MADEX
\citep{Cernicharo2012}. We have adopted the dipole moment, 3.49 D, predicted from the $ab$ $initio$ calculations
of \citet{Gronowski2013}. 
Four features in harmonic relation 8:10:12:13 have been found in IRC+10216 and are assigned to 
the rotational transitions J=8-7, 10-9, 12-11, and 13-12 of HMgNC. The lines are shown in 
Figure~\ref{fig:astronomical} and their frequencies are given in Table 1. Using our MADEX catalogue
we conclude that none of these lines agrees
in frequency with any isotopologue or vibrational level of already known molecular species.
From a fit to the observed frequencies we obtain $B_0$=5481.49(3)\,MHz and $D_0$=3.2(1)\,KHz, i.e., 
near identical to the rotational constants obtained in the laboratory.
The J=8-7 and J=10-9 transitions are
unblended with other features and are clearly detected. The J=9-8 line is predicted at 98657.3(2)\,MHz
and is fully blended with the the 11$_{1,10}$-10$_{1,9}$ line of H$_2$C$_4$ at 98655.094(4)\,MHz. Hence, it 
is not show in Figure~\ref{fig:astronomical}. The J=11-10 line at 120576\,MHz is fully blocked by the Earth atmosphere. 
The J=10-9 line of HMgNC at 109616.3 MHz agrees with the J=12-11 l=$\pm$2 transition of the $2\nu_6$ vibrational
level of HCCCN (1000 cm$^{-1}$ in energy). No lines of HCCCN $\nu_6=1,2$ are detected in our data from the
J=9-8 up to J=16-15 rotational transitions of these levels (see also \citet{Cernicharo2000}). 
The only vibrational level detected so far for HCCCN
is the $\nu_7$=1 at $\simeq$100 cm$^{-1}$. Hence, we are confident that this feature is unknown previous to its
assignment to HMgNC.
The J=12-11 transition of HMgNC at 131534.4(5)\, MHz is slightly contaminated in its red part
by a narrow, half power linewidth $\simeq$3\,MHz, and weak feature
of HCN in its $\nu_1+3\nu_2+\nu_3$ vibrational mode \citep{Cernicharo2011}.  
The CCS line in the high frequency part of the
line profile is separated by 17\,MHz from the J=12-11 transition of HMgNC and is not affecting the line profile.
Consequently, the HMgNC line is clearly detected with a signal to noise ratio (SNR) of $>$ 6.
Finally, the J=13-12 transition is predicted at 142491.8(7)\,MHz. It
is partially blended with the 11$_{11}$-10$_{10}$ line of CCS at 142501.694(6)\,MHz. At this frequency
the full linewidth of the IRC+10216 features is 13.8 MHz. As the HMgNC and CCS lines are separated by
10\,MHz 70\% of the HMgNC line profile is free of contamination by CCS (note that this spectrum has a sensitivity
of 0.6 mK). The U shaped profile of the
lines arising in the shell where radicals have been found is clearly visible in all the observed
lines. The rotational constants of HMgNC can be slightly improved by fitting the space frequencies together 
with the laboratory ones. We obtain $B_0$=5481.4336(4)\,MHz and $D_0$=2.94(5)\,KHz. Predictions for
lines with E$_{upp}$<100\,K are given in Table 1.

From the data provided in Figure~\ref{fig:astronomical} we obtain for HMgNC
T$_{rot}$=21$\pm$6\,K and $N$=(6$\pm$2)$\times$10$^{11}$\,cm$^{-2}$.
Using all the lines of MgNC and MgCN published in previous papers \citep{Guelin1986,Guelin1993,Ziurys1995} 
we derive for MgNC T$_{rot}$=18.6$\pm$1\,K and $N$=(13$\pm$3)$\times$10$^{12}$\,cm$^{-2}$.
For MgCN we obtain T$_{rot}$=13$\pm$3\,K and $N$=(7.4$\pm$2)$\times$10$^{11}$\,cm$^{-2}$.
The abundance ratio between these species is $N$(MgNC)/$N$(HMgNC)$\simeq$20, 
$N$(HMgNC)/$N$(MgCN)$\simeq$0.8, and $N$(MgNC)/$N$(MgCN) $\simeq$15.

\section{Chemical Modelling and Discussion}
The line profiles of HMgNC observed in IRC +10216 have a U shape which indicate that this molecule is formed in the 
outer shells, as the related radical MgNC \citep{Guelin1993}. The chemistry of HMgNC is highly uncertain 
--to our knowledge it has not been considered in the past-- although it is reasonable to think on the three Mg--bearing molecules 
detected in IRC +10216 (MgNC, MgCN, and HMgNC) sharing a common origin in the cold outer envelope. The formation of MgNC in 
circumstellar sources has been broadly discussed by \citet{Petrie1996} and \citet{Dunbar2002}, who favour a gas phase 
formation route driven by the radiative association of Mg$^+$ and cyanopolyynes

\begin{equation}
{\rm Mg^+} + {\rm HC}_{2n+1}{\rm N} \rightarrow {\rm Mg(HC}_{2n+1}{\rm N)}^+ +  h\nu, \label{reac:radiative-association}
\end{equation}

\noindent
followed by the dissociative recombination of the cation complex.
We have implemented the above chemical scheme for Mg in a chemical model of IRC +10216 similar to that by 
\citet{Agundez2008}. The rate constants for reaction (\ref{reac:radiative-association}) with $n$ = 3, 5, 7, and 
9 have been taken from the calculations by \citet{Dunbar2002}. The analogous reaction with HCN is too slow compared 
with those involving cyanopolyynes \citep{Petrie1996}. The dissociative recombination of the different cation complexes 
Mg(HC$_{2n+1}$N)$^+$ with free electrons may ultimately lead to Mg--containing cyanides. However, the branching ratios 
are unknown and are difficult to predict just based on the exothermicity of the reactions. 
If this mechanism indeed governs the 
formation of Mg--bearing molecules in IRC +10216 we should expect the main channel to yield MgNC, based on the relative 
observed abundances of MgNC, MgCN, and HMgNC. This latter species has a closed shell electronic structure and is thus 
expected to be less reactive than the two other radicals. We have therefore restricted the loss processes 
of HMgNC to just photodissociation by interstellar ultraviolet photons, while for MgNC and MgCN we have also considered 
reactions with H atoms and electrons. The absolute abundances of Mg--bearing molecules scale with the elemental abundance 
of magnesium in the gas phase, which unfortunately was not constrained by the observations of gas phase atoms 
in IRC +10216 carried out by \citet{Mauron2010}.

\begin{figure}
\includegraphics[angle=0,scale=.57]{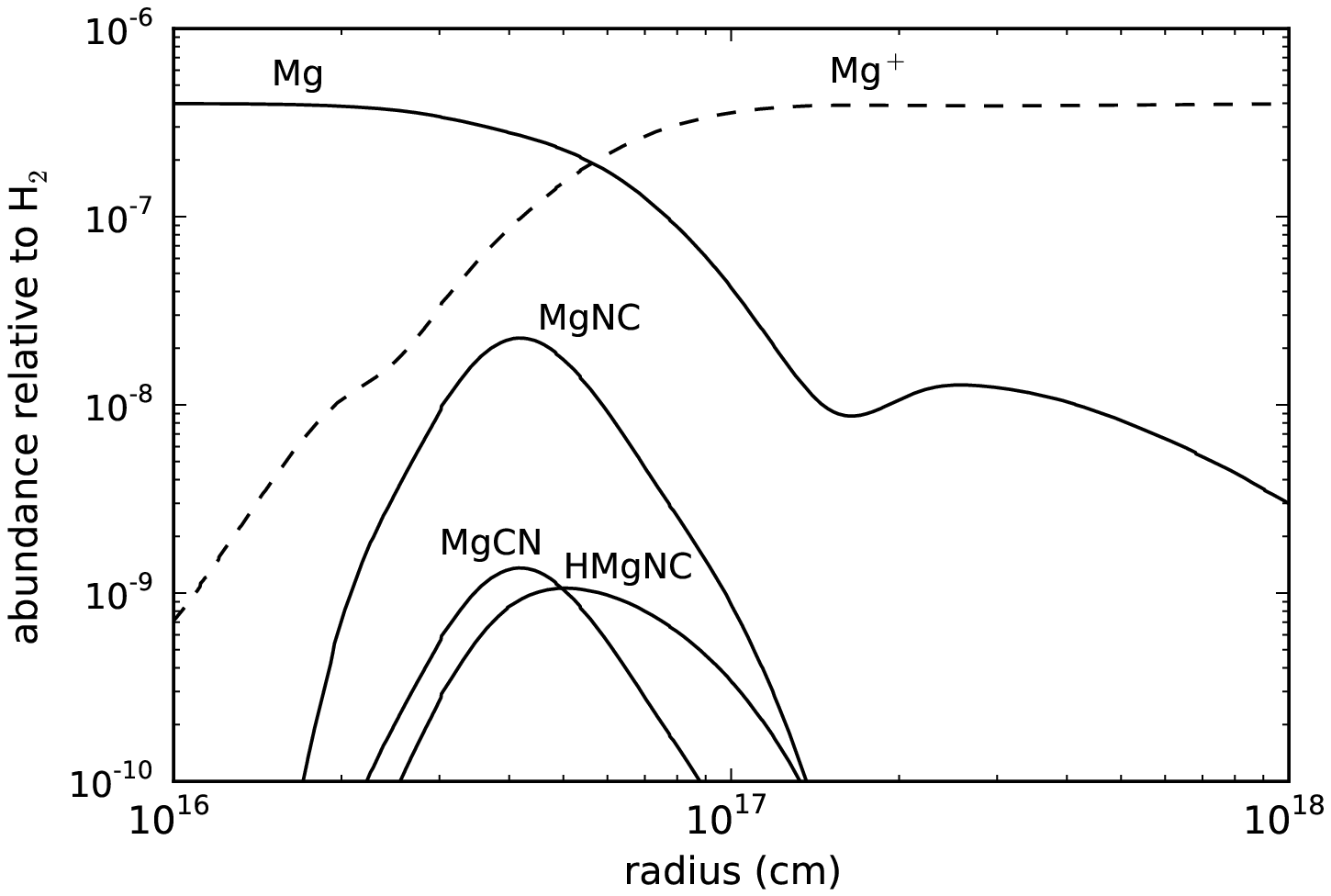}
\caption{Calculated abundances of Mg--bearing species in IRC +10216 as a function of the radial distance.}
\label{fig:chemistry}
\end{figure}

With this simple chemical scheme we find that in order to reproduce the relative and absolute column 
densities of the three Mg--bearing molecules observed in IRC +10216, we need a gas phase Mg abundance 
of 2$\times$10$^{-7}$ relative to H (i.e. depleted by a factor of $\sim$200 with respect to the solar abundance) 
and branching ratios in the dissociative recombination of Mg(HC$_{2n+1}$N)$^+$ ions, which for MgCN and HMgNC 
are 5 and 1\%, respectively, of that of MgNC. In Figure~\ref{fig:chemistry} we show the calculated 
abundances of Mg--bearing species, which are in good qualitative agreement with those obtained by \citet{Millar2008} 
on MgNC, except for some quantitative differences which are likely to arise from differences in the adopted chemical 
network. Similarly to \citet{Millar2008}, we find that the reactions of radiative association between Mg$^+$ and 
cyanopolyynes larger than HC$_3$N dominate the synthesis of Mg--bearing molecules, due to their higher rate constants.
The predicted abundance ratios for MgNC, HMgNC and MgCN are in good agreement with the observations. 
The putative common chemical origin of MgNC, MgCN, and HMgNC in IRC +10216 may be proven by mapping their emission 
distribution, which, except for some possible differences in the rotational excitation and chemical loss processes, 
should peak at similar angular distances. Investigation on the dissociative recombination of Mg(HC$_{2n+1}$N)$^+$ ions, 
particularly on the product channels and branching ratios, would also allow to shed light on the chemical synthesis 
of Mg--bearing molecules, and possibly of other metal--bearing cyanides and isocyanides which have been 
also detected in IRC +10216. Future observations with ALMA of metal (Mg, Ca, Fe)-bearing radicals 
and their corresponding hydrogenated closed shell molecules could provide important information on where
the different species are formed in the circumstellar envelope and, hence, on their formation mechanism.

\acknowledgments
The Spanish authors thank the Spanish MICINN for funding support through grants CTQ2010-19008, CSD2009-00038, 
AYA2009-07304, AYA2012-32032, and Junta de Castilla y Le\'on (grant VA070A08).

\clearpage


\begin{thebibliography}{}

\bibitem[Ag\'undez et al. (2008)]{Agundez2008} Ag\'undez, M., Fonfr\'ia, J. P., Cernicharo, J., et al. 2008, \aap, 479, 493
\bibitem[Alonso et al. (2009)]{Alonso2009}Alonso, J. L., P\'erez, C., Sanz, M. E., et al., 2009, Phys. Chem. Chem. Phys., 11, 617
\bibitem[Cabezas et al. (2012)]{Cabezas2012}Cabezas, C., Mata, S., Daly, A. M. et al., 2012, J. Mol. Spectrosc., 278, 31
\bibitem[Cernicharo (1985)]{Cernicharo1985} Cernicharo, J., 1985, Internal IRAM report (Granada: IRAM)
\bibitem[Cernicharo et al. (1986a)]{Cernicharo1986a}Cernicharo, J., Kahane, C., G\'omez-Gonz\'alez, J.,  \& Gu\'elin, M., 1986a, \aap, 164, L1
\bibitem[Cernicharo et al. (1986b)]{Cernicharo1986b}Cernicharo, J., Kahane, C., G\'omez-Gonz\'alez, J.,  \& Gu\'elin, M., 1986b, \aap, 167, L5
\bibitem[Cernicharo \& Gu\'elin (1987)]{Cernicharo1987} Cernicharo, J., Gu\'elin, M., 1987, \aap, 183, L10
\bibitem[Cernicharo et al. (1987a)]{Cernicharo1987a} Cernicharo, J., Gu\'elin, M., Walmsley, 1987a, \aap, 172, L5
\bibitem[Cernicharo et al. (1987b)]{Cernicharo1987b} Cernicharo, J., Gu\'elin, M., Menten, K.M., Walmsley, 1987b, \aap, 181, L1
\bibitem[Cernicharo \& Gu\'elin (1996)]{Cernicharo1996} Cernicharo, J., Gu\'elin, M., 1996, \aap, 309, L27
\bibitem[Cernicharo et al. (2000)]{Cernicharo2000} Cernicharo, J., Gu\'elin, M., Kahane, C., et al., 2000, A.\&A. SS, 142, 181
\bibitem[Cernicharo et al. (2011)]{Cernicharo2011} Cernicharo, J., Ag\'undez, M., Kahane, C., et al., 2011, \aap, 529, L3
\bibitem[Cernicharo (2012)]{Cernicharo2012}Cernicharo, J., 2012, in ECLA-2011: Proceedings of the European Conference 
on Laboratory Astrophysics, EAS Publications Series, vol 58, 2012, Editors: C. Stehl, C. Joblin, and L. d'Hendecourt (Cambridge: Cambridge Univ. Press), 251
\bibitem[Dunbar \& Petrie (2002)]{Dunbar2002} Dunbar, R. C. \& Petrie, S. 2002, \apj, 564, 792
\bibitem[Gronowski \& Kolos (2013)]{Gronowski2013} Gronowski, M., Kolos, R., 2013, J. Phys. Chem. A, 117, 4455
\bibitem[Gu\'elin et al. (1986)]{Guelin1986} Gu\'elin, M., G\'omez-Gonz\'alez, J., Cernicharo, J., Kahane, C., 1986, \aap, 157, L17
\bibitem[Gu\'elin et al. (1987)]{Guelin1987} Gu\'elin, M., Cernicharo, J., Kahane, C., G\'omez-Gonz\'alez, J., \& Walmsley, C.M., 1987, \aap, 175, L5
\bibitem[Gu\'elin, Lucas \& Cernicharo (1993)]{Guelin1993} Gu\'elin, M., Lucas, R., Cernicharo, J., 1993, \aap, 280, L19
\bibitem[Gu\'elin et al. (1997)]{Guelin1997} Gu\'elin, M., Cernicharo, J., Travers, M.J., et al., 1997, \aap, 317, L1
\bibitem[Gu\'elin et al. (2000)]{Guelin2000} Gu\'elin, M., Muller, S. Cernicharo, J., et al., 2000, \aap, 363, L9
\bibitem[Gu\'elin et al. (2004)]{Guelin2004} Gu\'elin, M., Muller, S. Cernicharo, J., et al., 2004, \aap, 426, L49
\bibitem[Highberger et al. (2003)]{Highberger2003} Highberger, J.L., Thomson, K.J., Young, P.A., et al., 2003, \apj, 593, 393
\bibitem[Kaminski et al. (2013)]{Kaminski2013} Kaminski, T., Gottlieb, C.A., Menten, K.M., et al., 2013, \aap, 551, A113.
\bibitem[Kawaguchi et al. (1993)]{Kawaguchi1993} Kawaguchi, K., Kagi, E., Hirano, T., et al., 1993, \apj, 406, L39
\bibitem[Mauron \& Huggins (2010)]{Mauron2010} Mauron, N., Huggins, P.J., 2010, \aap, 513, A31
\bibitem[Millar (2008)]{Millar2008} Millar, T. J. 2008, \apss, 313, 223
\bibitem[Pardo et al. (2001)]{Pardo2001} Pardo, J. R., Cernicharo, J., Serabyn, E. 2001, IEEE Trans. Antennas and Propagation, 49/12, 1683
\bibitem[Petrie (1996)]{Petrie1996} Petrie, S. 1996, \mnras, 282, 807
\bibitem[Pulliam et al. (2010)]{Pulliam2010} Pulliam, R.L., Savage, C., Ag\'undez, M., et al., 2010, \apj, 725, L181
\bibitem[Sanz et al. (2002)]{Sanz2002} Sanz, M.E., McCarthy M.C., Thaddeus, P., 2002, \apj, 577, L71
\bibitem[Tsuji (1973)]{Tsuji1973} Tsuji, T., 1973, \aap, 23, 411
\bibitem[Turner et al. (1994)]{Turner1994} Turner, B.E., Steimle, T.C., \& Meerts, L., 1994, \apj, 426, L97
\bibitem[Walker \& Gerry (1998)]{Walker1998}Walker, K. A. \& Gerry, M. C. L., 1998, J. Mol. Spectrosc., 189, 40
\bibitem[Zack et al. (2011)]{Zack2011} Zack, L.N., Halfen, D.T. \& Ziurys, L.M., 2011, \apj, 733, L36
\bibitem[Ziurys et al. (1995)]{Ziurys1995} Ziurys, L.M., Apponi, A.J., Gu\'elin, M., \& Cernicharo, J., 1995, \apj, 445, L47
\bibitem[Ziurys et al. (2002)]{Ziurys2002} Ziurys, L.M., Savage, C., Highberger, J.L., et al., 2002, \apj, 564, L45

\end{thebibliography}
\end{document}